# Spin Transport through the metallic antiferromagnet FeMn

H. Saglam[1,2,a], W. Zhang[1,4,b], M.B. Jungfleisch[1], J. Sklenar[1,3], J. E. Pearson[1], J. B. Ketterson[3] and A. Hoffmann[1,c]

[1] Materials Science Division, Argonne National Laboratory, Lemont IL 60439, USA

[2] Department of Physics, Illinois Institute of Technology, Chicago IL 60616, USA

[3] Department of Physics and Astronomy, Northwestern University, Evanston IL 60208, USA

[4] Department of Physics, Oakland University, Rochester MI 48309, USA

a) saglam@anl.gov
b) weizhang@oakland.edu
c) hoffmann@anl.gov

**Abstract**

We investigate spin transport through metallic antiferromagnets using measurements based on spin pumping combined with inverse spin Hall effects in $Ni_{80}Fe_{20}$/FeMn/W trilayers. The relatively large magnitude and opposite sign of spin Hall effects in W compared to FeMn enable an unambiguous detection of spin currents transmitted through the entire FeMn layer thickness. Using this approach we can detect two distinctively different spin transport regimes, which we associate with electronic and magnonic spin currents respectively. The latter can extend to relatively large distances ($\approx$ 9 nm) and is enhanced when the antiferromagnetic ordering temperature is close to the measurement temperature.

The past three decades have seen an amazing transformation of the spintronics research field accompanied by an impressively fast transition of fundamental physics research into applications revolutionizing information technologies.[1] The first decade, inspired by the discovery of giant magnetoresistance,[2,3] focused on the interplay between magnetization structure and charge transport mediated by spin polarized charge currents and had direct impact on magnetic data storage. During the second decade, pure spin currents, generated and detected via non-local measurements[4,5,6,7] or spin Hall effects,[8,9,10] gained increased interest and set the foundation for spin-orbitronics concepts leading to the incorporation of magnetic phenomena into logic and data processing.[1] More recently, it has been realized that magnetization dynamics itself can be the source and conduit for spin currents, which generated a new research direction called magnonics,[11] where spin information is transmitted via the fundamental quasiparticle excitations of magnetically ordered systems, magnons or spin waves.

An interesting new addition to this wide variety of phenomena is the discovery that antiferromagnetic materials may play an active role in spintronic devices.[12,13] Traditionally, antiferromagnets have been used to provide magnetic reference orientation in magnetic read heads and magnetic memory elements through the exchange bias mechanism,[14,15] which describes a magnetic unidirectional anisotropy that develops due to the magnetic interaction between an antiferromagnet and a ferromagnet. However, more recently it has been recognized that non-collinear antiferromagnetic spin structures result in geometric Berry-phases that profoundly change charge transport by generating large anomalous Hall effects.[16,17,18] Furthermore, large spin Hall effects have been measured in metallic antiferromagnet[19] with spin Hall conductivities sufficiently large to enable the manipulation of magnetization of an adjacent ferromagnet.[20] This already gave rise to the development of novel memristive devices.[21] Even more interestingly, it has recently been recognized that spin-orbit torques in conducting antiferromagnets may have the same symmetry as the staggered ordered antiferromagnetic spin structure, and thus can provide a direct way to electrically manipulate antiferromagnetic order.[22]

All these investigations have focused on metallic antiferromagnets. However, even insulating antiferromagnets can be active components of spin transport devices. Towards this end it has been shown that heat-currents can generate spin currents via the spin Seebeck effect[23,24] in a way that is similar to previous measurements based on ferromagnetic insulators.[25] Furthermore, it has been recognized that spin currents can be effectively transmitted through insulating antiferromagnets[26,27] and can even be amplified at critical thickness.[28,29] A theoretical proposal to explain such amplification suggests that the spin transport occurs via evanescent spin waves, which extract additional angular momentum from the lattice of the antiferromagnet.[30] Alternatively it has also been suggested that the spin current transmission can be understood in terms of magnon diffusion.[31] At the same time it appears that the temperature dependence of these spin transport phenomena are closely correlated with the ordering temperature[32,33] and are reminiscent of the magnetic susceptibility in thin antiferromagnetic films.[34] Thus these spin currents in antiferromagnetic insulators appear to be closely related to spin fluctuation, which become strongly enhanced at the Neél temperature.

The natural question occurs, whether similar spin fluctuation driven spin currents also occur in metallic antiferromagnets. We investigated this possibility in Py/FeMn/W (Py = $Ni_{80}Fe_{20}$) trilayers using the widely used metallic antiferromagnet FeMn and spin pumping–inverse spin Hall effect measurements. Compared to previous spin pumping-inverse spin Hall effect measurements,[19,20] the additional W layer enables to detect, whether a spin current traverses the entire FeMn layer thickness. This is possible, since the spin Hall angle of W ($\Theta_{SH}$ = -0.33)[10,35] has an opposite sign compared to FeMn ($\Theta_{SH}$ = +0.008),[19,20] and is one order of magnitude larger. Thus the sign of the spin Hall voltage signal provides an unambiguous indication of spin currents reaching the W layer. As will be discussed in detail below we detect as a function of FeMn thickness a strongly non-monotonic dependence of the spin Hall voltage, which suggests that there are two clearly different regimes of spin current transport in FeMn. We assign these two regimes to spin currents carried by electrons and magnons, respectively.

The sample preparation was analogous to previous experiments;[36] namely, we defined on top of a Si substrate with 300-nm thick thermally grown $SiO_2$ layer rectangular samples of Py/FeMn/W trilayers with lateral dimensions of 20 μm×200 μm using photolithography, magnetron sputter deposition of the three layers, and subsequent lift-off. For all samples the thickness of the Py and W layers were kept constant at 12 and 4 nm, respectively, while the thickness of the intermediate antiferromagnetic FeMn layer, $t_{FeMn}$, was varied from 0 to 12 nm. Finally, electric leads made of Ti(3 nm)/Au(150 nm) were added to the trilayer before it was covered first by 80-nm thick insulating MgO layer and a coplanar waveguide [see Fig. 1(a)].

All transport measurements presented here were performed at room temperature. The magnetization dynamics was excited by applying a 10 mW microwave current to the coplanar waveguide with frequencies ranging from 4 to 10 GHz. The precessional motion of the magnetization in the Py layer leads to the injection of angular momentum into the FeMn, a phenomenon known as spin pumping.[37] This injected spin current, which is perpendicular to the film interfaces, is then converted into a measurable transverse electric voltage via the inverse spin Hall effect.[38] This transverse voltage is generated in a direction perpendicular to the equilibrium magnetization direction in the Py layer, which is given by the externally applied magnetic field. The magnetic field is applied at a 40° angle with respect to the coplanar waveguide in order to enable both efficient excitation of the magnetization dynamics and the measurements of net voltages along the long axis of the samples. Also note that the inverse spin Hall voltages in the FeMn and W layers have opposite signs due to the opposite sign of the spin Hall angle in these two materials.[10,19,20,35]

Typical measured spectra for different FeMn layer thicknesses $t_{FeMn}$ are shown in Fig. 2 for measurements performed with microwave excitation at 4 GHz. The measured dc voltage spectra as a function of field show a superposition of symmetric and antisymmetric Lorentzian components. It was shown previously that for our experimental geometry[39,40] these two different components can be attributed to the inverse spin Hall effect (symmetric) and homodyne anisotropic magnetoresistance (antisymmetric) due to mixing of rf currents. In the previous measurement, the spin Hall effect contribution to the measurement voltage signals increased monotonically and saturated with the thickness of the layer generating the spin Hall voltage[19,20]. In contrast to that, the data shows an unusual trend as a function of $t_{FeMn}$. Without any FeMn layer, *i.e.*, with the Py in direct contact with W, the symmetric Lorentzian voltage contribution is negative, as is expected from the negative sign of the spin Hall angle for W.[10,35] Then, by adding increasingly thicker FeMn layers the symmetric Lorentzian contribution associated with the spin Hall effect is initially reduced. This can be easily understood in that the spin Hall voltage generated in FeMn is opposite to the spin Hall voltage generated in W and at the same time due to the relatively short spin diffusion length in FeMn of 1–2 nm,[19, 41] one expects the spin current reaching the W layer to be reduced. However, upon further increasing $t_{FeMn}$, the voltage signal due to the spin Hall effects grows again in magnitude with a negative sign. Interestingly, for $t_{FeMn}$ = 8 nm the spin Hall effect voltage even exceeds the voltage measured without FeMn, implying that for this thickness of FeMn an even larger spin current is injected into the W layer than without any FeMn layer. Note that one would expect additional current shunting and spin Hall voltages with opposite sign from the FeMn layer. Upon further increasing $t_{FeMn}$ to 10 nm, the sign of the spin Hall voltage switches, indicating that now the spin Hall voltage is dominated by spin currents in the FeMn layer and that at best a relatively small spin current reaches the W layer.

In order to better illustrate this dependence of the inverse spin Hall voltage on the thickness of the FeMn layer, we can define the relative contribution of the spin Hall voltage to the total voltage as[19,20,36] $W_{ISHE} = V_{ISHE}/(|V_{ISHE}|+V_{AMR})$, where $V_{ISHE}$ and $V_{AMR}$ are the amplitudes of the symmetric and antisymmetric contributions to the measured voltage spectra. Since all voltages scale linear with the applied microwave power, this definition avoids variations in measured absolute voltages due to inadvertent power variations for measurements with different samples. The resulting values are shown in Fig. 3 for measurements on samples with $t_{FeMn}$ varying between 0–12 nm, and applied microwave frequencies varying from 4–9 GHz. As can be seen the same general trend is observed, independent of the measurement frequency. The observation of larger relative spin Hall signals for larger frequencies is related to a changing ellipticity of the ferromagnetic precession and is consistent with previous measurements analyzed similarly.[19,42] However, the non-monotonic behavior of $W_{ISHE}$ is in striking contrast to previous measurements, especially compared to those of Py/FeMn bilayers,[19] which only show a monotonic increase of $W_{ISHE}$ as a function of $t_{FeMn}$ and saturates above 6 nm. Instead we see an initial decrease of the inverse spin Hall signal followed by a gradual increase at 8 nm and a subsequent abrupt change in sign of the spin Hall voltage.

The observed thickness dependence correlates well with the well-known fact that exchange bias at room temperature is suppressed for thickness of FeMn below 8 nm, since for smaller thicknesses the antiferromagnetic order in FeMn is either not established or the anisotropy in FeMn is insufficient to stabilize a unidirectional anisotropy in the adjacent

Py.[41,43] The phase transition at 8 nm was further corroborated by the ferromagnetic resonance data extracted from the spin-pumping measurements of our samples. Figure 4 shows the resonance linewidth and field as a function $t_{FeMn}$. As $t_{FeMn}$ is increased the resonance field drops above 8 nm, which can be understood as the change of the effective field due to the additional uniaxial anisotropy, when exchange bias is established. Similarly, the linewidth gradually increases and shows a maximum above 8 nm, again coinciding with the onset of exchange bias. These observations are consistent with previous ferromagnetic resonance measurements on the Py/FeMn exchange bias system as a function of FeMn thickness.[44] Therefore, these results support the explanation that the large spin currents observed for FeMn thicknesses well above the electronic spin diffusion length are mediated by spin fluctuations, when the measurement temperature is above the thickness-dependent blocking temperature of the FeMn. This explanation is similar to a recent theoretical model that predicts spin pumping into a ferromagnetic material to be enhanced near the Curie temperature $T_c$ due spin fluctuations.[45] Indeed, measurements at 150 K show qualitatively similar behavior to the room temperature data, but now the sign change occurs for approximately 2 nm reduced FeMn thicknesses, consistent with a lower blocking temperature for thinner FeMn[43]. Furthermore, we notice that the length-scale for these magnonic spin currents is comparable to previous measurements on the insulating antiferromagnet NiO.[26,28,29]

Using spin pumping and inverse spin Hall effect measurements, we explored the spin current transmission through metallic antiferromagnet FeMn sandwiched in between Py and W. The large negative spin Hall angle of W, $\Theta_{SH}$ = -0.33,[10,35] enables unambiguous detection of spin current transmission even in the presence of spin Hall voltages from FeMn, which has a smaller positive spin Hall angle, ($\Theta_{SH}$ = +0.008).[19,20] As a function of the FeMn thickness we observe a non-monotonic dependence of the total spin Hall voltage. For small FeMn thicknesses, < 2 nm, there is a decreasing voltage signal, but above 2 nm the voltage signal increases in magnitude with a maximum for 8-nm thick FeMn, indicating that significant spin currents reach the W layer even for thicknesses well above the spin diffusion length in FeMn. At the same time for 8-nm thick FeMn the blocking temperature of exchange bias is close to the measurement temperature (300 K), which suggest that similar to insulating antiferromagnets the spin current is mediated by magnon excitations. The presence of two different spin current transport regimes due to either electronic or magnonic excitations may provide new concepts for integrating antiferromagnetic materials into spintronic devices.

This work was supported by the U.S. Department of Energy, Office of Science, Basic Energy Sciences, Materials Science and Engineering Division. Lithographic patterning was carried out at the Center for Nanoscale Materials, which is supported by the DOE, Office of Science, Basic Energy Sciences under Contract No. DE-AC02-06CH11357.

**Figures**

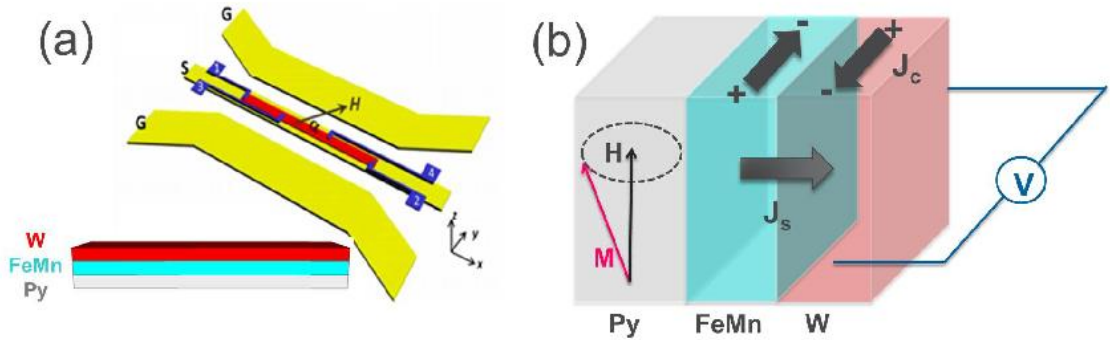

**Figure 1**. (a) Schematic diagram of the device with coplanar waveguide. (b) Schematic of spin pumping and the inverse spin Hall effect measurement on Py/FeMn/W trilayers.

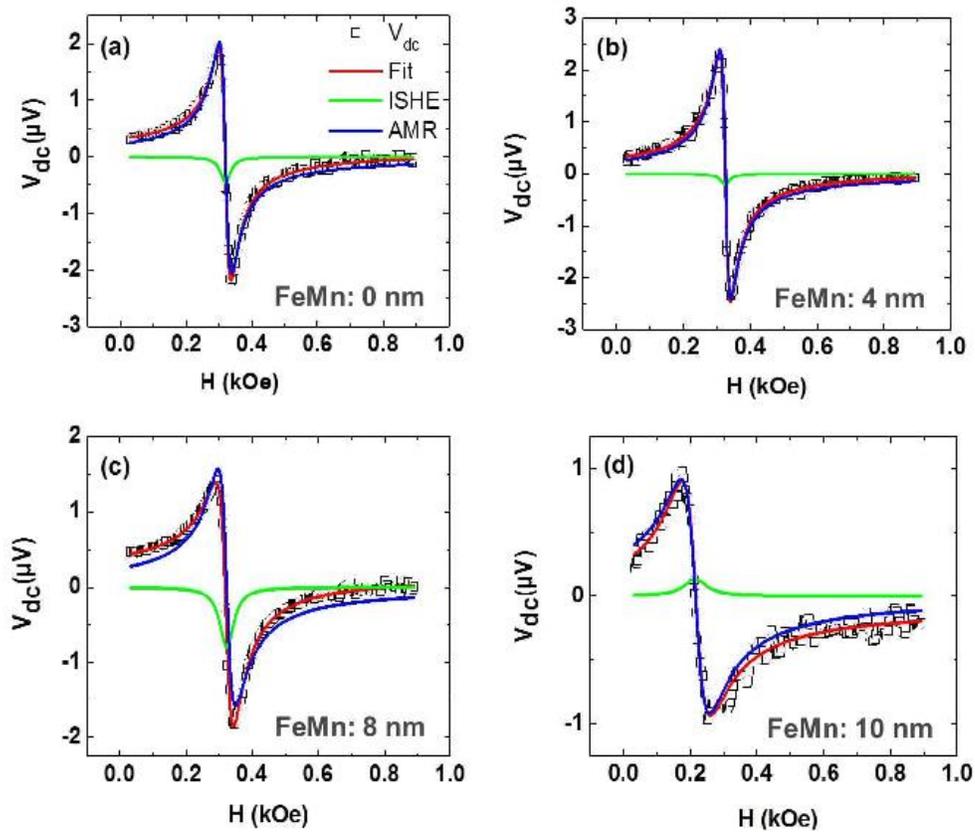

**Figure 2.** Anisotropic magnetoresistance – inverse spin Hall effect spectra measured at 4 GHz for Py(12)/FeMn($t$)/W(4) structures at room temperature for four different

thicknesses *t* of the FeMn layer; (a) 0 nm, (b) 4 nm, (c) 8 nm, and (d) 10 nm. Green and blue lines indicate the ISHE and AMR contributions to the combined fit shown in red. Black symbols represent experimental data.

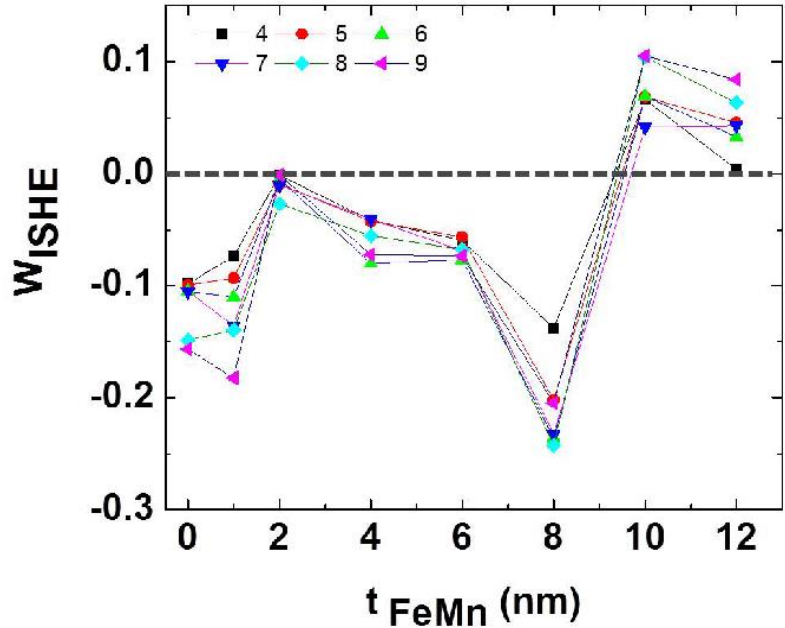

**Figure 3**. Thickness dependence of the relative spin Hall effect contribution to the total voltage $W_{ISHE}$ measured at room temperature for frequencies from 4 to 9 GHz.

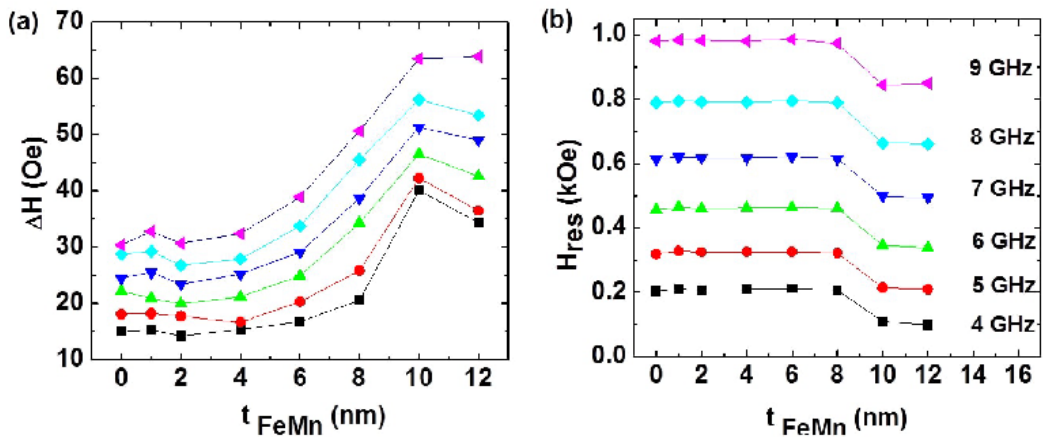

**Figure 4.** (a) FeMn thickness *t* dependence of the resonance linewidth $\Delta H$ measured at frequencies 4–9 GHz. (b) FeMn thickness *t* dependence of the resonance field $H_{res}$ measured at frequencies 4–9 GHz.

[1] A. Hoffmann and S. D. Bader, Phys. Rev. Appl. **4**, 047001 (2015).
[2] M. N. Baibich, J. M. Broto, A. Fert, F. N. van Dau, F. Petroff, P. Etienne, G. Creuzet, A. Friedrich, and J. Chazelas, Phys. Rev. Lett. **61**, 2472 (1988).
[3] P. Binasch, P. Grünberg, F. Saurenbach, and W. Zinn, Phys. Rev. B **39**, 4828 (1989).
[4] M. Johnson and R. H. Silsbee, Phys. Rev. Lett. **55**, 1790 (1985).
[5] F. J. Jedema, A. T. Filip, and B. J. van Wees, Nature **410**, 345 (2001).
[6] S. O. Valenzuela, and M. Tinkham, Nature **442**, 176 (2006).
[7] T. Kimura, Y. Otani, T. Sato, S. Takahashi, and S. Maekawa, Phys. Rev. Lett. **98**, 156601 (2007).
[8] E. Saitoh, M. Ueda, H. Miyajima and G. Tatara, Appl. Phys. Lett. **88**, 182509 (2006).
[9] O. Mosendz, J. E. Pearson, F.Y. Fradin, G. W. Bauer, S. D. Bader and A. Hoffmann, Phys. Rev. Lett. **104**, 046601 (2010).
[10] A. Hoffmann, IEEE Trans. Magn. **49**, 5172 (2013).
[11] A. V. Chumak, V. I. Vasyuchka, A. A. Serga, and B. Hillebrands, Nat. Phys. **11**, 453 (2015).
[12] T. Jungwirth, X. Marti, P. Wadley, and J. Wunderlich, Nat. Nanotechn. **11**, 231 (2016).
[13] J. Sklenar, W. Zhang, M. B. Jungfleisch, W. Jiang, H. Saglam, J. E. Pearson, J. B. Ketterson, and A. Hoffmann, AIP Adv. **6**, 055603 (2016).
[14] W. Zhang and K. M. Krishnan, Mater. Sci. Engr. R: Rep. **105**, 1 (2016).
[15] J. Nogués, J. Sort, V. Skumryev, S. Suriñach, J. S. Muñoz, and M. D. Baró, Phys. Rep. **422**, 65 (2005).
[16] R. Shindou and N. Nagaosa, Phys. Rev. Lett. **87**, 116801 (2001).
[17] I. Martin and C. D. Batista, Phys. Rev. Lett. **101**, 156402 (2008).
[18] S. Nakatsuji, N. Kiyohara, and T. Higo, Nature **527**, 212 (2015).
[19] W. Zhang, M. B. Jungfleisch, W. Jiang, J. E. Pearson, A. Hoffmann, F. Freimuth, and Y. Mokrousov, Phys. Rev. Lett. **113**, 196602 (2014).
[20] W. Zhang, M. B. Jungfleisch, F. Freimuth, W. Jiang, J. Sklenar, J. E. Pearson, J. B. Ketterson, Y. Mokrousov, and A. Hoffmann, Phys. Rev. B **92**, 144405 (2015).
[21] S. Fukami, X. Zhang, S. DuttaGupta, A. Kuenkov, and H. Ohno, Nat. Mater. **15**, 535 (2016).
[22] P. Wadley, *et al.*, Science **351**, 587 (2016).
[23] S. Seki, T. Ideue, M. Kubota, Y. Kozuka, R. Takagi, M. Nakamura, Y. Kaneo, M. Kawasaki, and Y. Tokura, Phys. Rev. Lett. **115**, 266601 (2015).
[24] S M. Wu, W. Zhang, A. KC, P. Borisov, J. E. Pearson, J. S. Jiang, D. Lederman, A. Hoffmann, and A. Bhattacharya, Phys. Rev. Lett. **116**, 097204 (2016).
[25] K. Uchida, H. Adachi, T. Ota, H. Nakayam, S. Maekawa, and E. Saitoh, Appl. Phys. Lett. **97**, 172505 (2010).
[26] C. Hahn, G. de Loubens, V. V. Naletov, J. B. Youssef, O. Klein, and M. Viret, Europhys. Lett. **108**, 57005 (2014).
[27] T. Moriyama, S. Takei, M. Nagata, Y. Yoshimura, N. Matsuzaki, T. Terashima, Y. Tserkovnyak, and T. Ono, Appl. Phys. Lett. **106**, 162406 (2016).
[28] H. Wang, C. Du, P. C. Hammel, and F. Yang, Phys. Rev. Lett. **113**, 097202 (2014).
[29] W. Lin, K. Chen, S. Zhang, and C. L. Chien, Phys. Rev. Lett. **116**, 186601 (2016).

[30] R. Khymyn, I. Lisenkov, V. S. Tiberkevich, A. N. Slavin, and B. A. Ivanov, Phys. Rev. B **93**, 224421 (2016).
[31] S. M. Rezende, R. L. Rodríguez-Suárez, and A. Azevedo, Phys. Rev. B **93**, 054412 (2016).
[32] L. Frangou, S. Oyarzún, A. Auffret, L. Vila, S. Gambarelli, and V. Baltz, Phys. Rev. Lett. **116**, 077203 (2016).
[33] Z. Qiu, J. Li, D. Hou, E. Arenholz, A. T. N'Diaye, A. Tan, K.-i. Uchida, K. Sato, S. Okamoto, Y. Tserkovnyak, Z. Q. Qiu, and E. Saitoh, Nat. Comm. **7**, 12670 (2016).
[34] T. Abrose and C. L. Chien, Phys. Rev. Lett. **76**, 1743 (1996).
[35] C. F. Pai, L. Liu, H. W. Tseng, D. C. Ralph, and R. A. Buhrman, Appl. Phys. Lett. **101**, 122404 (2012).
[36] W. Zhang, M. B. Jungfleisch, W. Jiang, J. Sklenar, F. Y. Fradin, J. E. Pearson, J. B. Ketterson, and A. Hoffmann, J. Appl. Phys. **117**, 172610 (2015).
[37] Y. Tserkovnyak, A. Brataas, and G. E. W. Bauer, Phys. Rev. Lett. **88**, 117601 (2002).
[38] M. B. Jungfleisch, W. Zhang, W. Jiang, A. Hoffmann, SPIN **5**, 1530005 (2015).
[39] O. Mosendz et al., Phys. Rev. Lett. **104**, 046601 (2010).
[40] O. Mosendz et al., Phys. Rev. B **82**, 214403 (2010).
[41] Y. Yang, Y. Xu, X. Zhang, Y. Wang, S. Zhang, R.-W. Li, M. S. Mireshekarloo, K. Yao, and Y. Wu, Phys. Rev. B **93**, 094402 (2016).
[42] V. Vlaminck, J. E. Pearson, S. D. Bader, and A. Hoffmann, Phys. Rev. B **88**, 064414 (2013).
[43] H. Sang, Y. W. Du, and C. L. Chien, J. Appl. Phys. **85**, 4931 (1999).
[44] W. Stoecklien, S. S. P. Parkin, and J. C. Scott, Phys. Rev. B **38**, 6847 (1988).
[45] Y. Ohnuma, H. Adachi, E. Saitoh, and S. Maekawa, Phys. Rev. B **89**, 174417 (2014).